\documentclass{article}
\usepackage[utf8]{inputenc}
\usepackage{graphicx}
\usepackage[a4paper=true]{hyperref}
\usepackage{enumitem}
\usepackage[square,sort,comma,numbers]{natbib}
\usepackage{breakurl}

\usepackage{amsmath}
\usepackage{graphicx}

\renewcommand{\theenumii}{\theenumi.\arabic{enumii}.}   

\title{Modification of the Shower Database\\ of the IAU~Meteor Data Center}
\author{M\'{a}ria Hajdukov\'{a}$^1$, Regina Rudawska$^2$, Tadeusz J. Jopek$^3$,\\ Masahiro Koseki$^4$, Gulchehra Kokhirova$^5$, Lubo\v{s} Neslu\v{s}an$^6$}
\date{\small{$^1$Astronomical Institute, Slovak Academy of Sciences, Slovakia, $^2$RHEA group/ESA ESTEC, The Netherlands, $^3$Astronomical Observatory Institute, Faculty of Physics,  A.M. University, Poznań, Poland, $^4$The Nippon Meteor Society, Japan, $^5$Institute of Astrophysics, National Academy of Sciences of Tajikistan, Republic of Tajikistan, $^6$Astronomical Institute of the Slovak Academy of Sciences, Tatranská Lomnica, Slovak Republic}}

\begin{document}
\maketitle
\noindent \small{Accepted: Astronomy and Astrophysics}

\subsection*{\centering Abstract}
The Shower Database (SD) of the Meteor Data Center (MDC) has been operating for 15 years and is used by the entire community of meteor astronomers. It contains meteor showers categorised in individual lists on the basis of their status. Since the inception of the SD, no objective rules for moving showers between individual lists have been established. The content of the SD has not yet been checked for the correctness of the meteor data contained therein.

Our aims are (1) to formulate  criteria for nominating meteor showers for established status, (2) to improve the rules for the removal of showers, (3) to verify and enhance the content of the SD, and (4) to improve the user area of the MDC SD.

The criteria for moving showers from the Working list to the Lists of established or removed Showers 
were generated using an empirical evaluation of their impact on the registered showers. The correctness of the parameters of each stream included in the SD was checked by comparing them with the values given in the source publications.

We developed a set of criteria for nominating showers to be established. We objectified rules for the temporary and permanent removal of meteor showers from the Working list. Both of our proposed new procedures were approved by a vote of the commission F1 of the IAU in July 2022. 
   
We verified more than $1350$ data records of the MDC SD and introduced  $\sim1700$ corrections. We included new parameters for shower characterisation. As a result of our verification procedure, $117$ showers have been moved to the List of removed showers. As of October 2022, the SD contains $923$ showers, $110$ of which are in the List of established Showers and $813$ are in the Working list. We also improved the user area of the SD and added a simple tool to allow a quick check of the similarity of a new shower to those in the database.

\section{Introduction}
%
%
In astronomy, observational databases of celestial bodies (galaxies, stars, planets, comets, asteroids, etc.) form a foundation used to understand the world around us. It is no different when it comes to the meteoroid component of our Solar System. Thus, the creation and maintenance of a high-quality meteor shower database has an important role for the community of professional, as well as amateur, meteor astronomers.    

The collection of meteor shower data started at the Meteor Data Center (MDC) in 2007. Initially, the online database contained data on $230$ streams published in professional or amateur scientific journals. Over the years, the number of recorded streams has increased significantly, today reaching $900$, which is undoubtedly a result of advances in observational technology as well as growing interest in meteor topics. However, as is undoubtedly the case with any database, imperfections of various natures were noted in the MDC. These imperfections range from typographical errors to erroneous literature references to the sources from which the data were taken.

Furthermore, we found that, in the MDC, it is possible to encounter single streams, so-called duplicates, which are successive solutions of streams already existing in the database. 
On the contrary to these duplicates, there are also several records that relate to one shower but are in fact solutions of separate meteoroid streams (false-positive duplicates). All these drawbacks can lead to incorrect results when using the database.  
Therefore, we decided to revise the content of the MDC database, and the first results of this task are presented in the following chapters of this paper. In addition, we propose new procedures to facilitate the upkeep of the MDC database.  

In Section \ref{MDCbegi}, we give a brief historical overview of the MDC database, and in Section \ref{MDCstructure} we introduce the MDC shower database structure and present new submission rules. Section \ref{MDCestablished} answers the problem of nominating meteor showers for so-called established status and the clarification of such a status. In Section \ref{MDCremoved}, we describe problems with shower data removals and propose rules that are objective. The last two sections, \ref{MDCverification} and \ref{MDCnew}, are devoted to the verification process of the MDC database content and to its new version, including the new website interface. 

\section{Beginning of the MDC SD}
\label{MDCbegi}
   The MDC\footnote{
   \url{https://www.iaumeteordatacenter.org/}.} 
   operates at the Astronomical Institute of the Slovak Academy of Sciences under the auspices of the International Astronomical Union (IAU). 
   The MDC has two parts: the database of individual Meteoroid Orbits (MO, see  \citet{1987PAICz..67..201L}, \citet{2003EMP...93..249L}, and \citet{2014EMP..111..105N}), and the Shower Database (SD), which the present paper is concerned with. The development of the SD and  
   its regulation and management rules, can be found in these publications: \citet{2010IAUTB..27..158B}, \citet{2010IAUTB..27..177W}, \citet{2011msss.conf....7J}, \citet{2014me13.conf..353J}, \citet{2017PSS..143....3J}, \citet{2020PSS..18204821J} and \citet{2021JIMO...49..163R}.
   
   The SD of the MDC is responsible for the unique designation of each new meteor shower and serves as a central register of meteor shower names and their basic parameters for the whole meteor science community. The SD acts in conjunction with the Working Group (WG) on Meteor Shower Nomenclature of the IAU Commission F1, 'Meteors, Meteorites, and Interplanetary Dust'.
   
   At the 2006 IAU general assembly in Prague, on the initiative of Peter Jenniskens, Commission 22 established the Task Group for Meteor Shower Nomenclature
   with the objective of formulating a list of meteor showers that can receive official names during the 2009 IAU general assembly in Rio de Janeiro, \citep[see][]{2007IAUTB..26..140S}.\footnote{In 2009 the Task Group was renamed the Working Group on Meteor Shower Nomenclature.}
In 2007, the SD 
was created as an independent part of the IAU MDC and was posted on the website\footnote{ \url{https://www.ta3.sk/IAUC22DB/MDC2007/}}. In the beginning, selected parameters of about $230$ meteor showers from the book \citep{2006mspc.book.....J}, kindly provided to the MDC by the author, were included in the database.

It is worth pointing out that the creation of the MDC SD was not intended to collect all of the parameters related to a given meteor shower. Its purpose was to give each new meteoroid stream, before publication in the scientific literature, a unique name and codes. To achieve this, it required codification of the rules used in the scientific literature for the naming of meteoroid streams or meteor showers.   
   
   To date (October 2022), $1040$ showers have been registered at the IAU MDC SD, of which the names of $110$ have been officially approved by the IAU (\citet{2011msss.conf....7J}, \citet{2014me13.conf..353J}, and \citet{2017PSS..143....3J}). More than $800$ showers have yet to be confirmed. $117$ showers have been disproven or found to be duplicates of a meteor shower detected previously. The current status of the SD, with showers categorised in the individual lists, is shown in Table \ref{tab:msdlists}.

   To eliminate deficiencies in the categorisation of meteor showers  due to the lack of objective rules, here we introduce changes to various procedures and propose criteria for the maintenance of individual lists of meteor showers. To fix inconsistent types of shower data and other failings within the SD, we verified all the records in the SD.

\section{Database structure and submission procedure}
\label{MDCstructure}
\subsection{The lists of meteor showers}
\label{structure}
About $1000$ meteor showers are stored in the MDC database.  
Showers have a different status and have been grouped into four lists. The content of these lists may be displayed by the web browser and downloaded as ASCII files.

{\it{The List of all showers}} contains all showers that are registered in the SD, except for those that are in the List of removed showers. 

{\it The List of established showers} is a subset of the List of all showers. It contains showers whose names were officially approved during the IAU general assembly. These showers have been given 'established status'.

{\it The Working list} (also a subset of the List of all showers) contains showers submitted to the MDC that have not yet been confirmed ('working' status). They have to be (within 6 months after submission) published in the scientific literature or in journals dedicated to amateur astronomers, WGN (the Journal of the IMO) and Meteor News. 
The shower remains in the Working list until it meets the criteria for established status (see Sect. \ref{C-criteria}) or is demoted to the List of removed showers (see Sect. \ref{R-criteria}).

{\it The List of removed showers} 
contains showers that have been removed from the Working list for various  reasons. These showers have been given status 'removed'. \\

There used to be a fifth list in the SD, the {\it List of shower groups} ('groups' status), which contained shower groups or complexes suggested in scientific publications. We decided to remove this list from the MDC because of the many difficulties and/or inconsistencies connected with this category. The category of stream groups was removed from the MDC and their members were moved to the Working list.

The mean characteristics of some showers were determined by more than a single author team. Hence, the meteor showers in the SD may be represented by two or more sets of radiants and orbital parameters. Hereafter, we speak of these multiple sets of parameters as 'multiple solutions' (or simply as 'solutions'). 
The solutions have significant functions aside from their cognitive importance. Each independent set of parameters strongly confirms the existence of a particular meteor shower. Despite this fact, there are only $195$ showers with more than one solution in the MDC SD (see Section \ref{R-criteria}). 
We encourage observers to also submit shower data of known meteor showers, not just those of newly discovered showers.

\begin{center}
\begin{table}[!ht]
\centering
\caption{Lists of showers included in the MDC, status as of October 2022. }
\begin{tabular}{lrr}
\hline
\centering
{Type of list} &   {Number of} & {Number of}  \\
{} &   {meteor showers} & {records (solutions)}   \\
\hline
{List of All Showers} & 923 & 1325\\
{List of Established Showers} & 110 & 368\\
{Working List} &  813 & 957   \\
{List of Removed Showers} &  117 & 208 \\
\hline
\end{tabular}
\label{tab:msdlists}
\end{table}
\end{center}

\subsection{Data submission procedure}
\label{procedure}
Since the very beginning of the MDC SD, each new meteoroid stream (meteor shower) has been given a unique name, and both a numerical and a three-letter code, at the moment of providing the shower data to the MDC. 
The discoverer providing the data of a meteor shower was asked to propose a new, unique name, according to the shower nomenclature rules. These, however, were recently transformed from this old one-step process for determining the final names of showers to a new two-step procedure. The new nomenclature rules are described in detail in a separate paper by \citet{2022AA...02..1H}. 

The observer has to provide the meteor shower data in the required formats (see the next section). Then, the submitted data of the new stream are placed on the Working list of the MDC SD. Subsequently, the discoverer of the stream has six months to provide a copy of the publication to the MDC, describing their discovery. The MDC SD team accepts meteor showers published in all peer-reviewed scientific journals in the field of astronomy, as well as in the two journals dedicated to amateur meteor astronomy: WGN, the Journal of the IMO, and Meteor News. If the publication does not reach the MDC on time (within 6 months of the submission date), the stream codes, name, and parameters are permanently removed from the database.  

\subsection{Submission rules}
\label{submission}

\subsubsection{Shower mean data}
When submitting, authors are asked to provide shower data in two files using the templates from the MDC website: the shower mean data and the look-up table.
The shower mean data file contains the averaged geocentric and orbital parameters, as well as additional information. Some data are mandatory (marked with an *), while some are optional. \\

The data are as follows: 
submission date UTC - date of the data submission to MDC, format YYYY.MM.DD*; 
ecliptic longitude of the Sun at the beginning and the end of the shower activity, [deg], J2000*;
ecliptic longitude of the Sun $\lambda_S$ corresponding to the maximum of the shower activity, [deg], J2000*;
geocentric right ascension $\alpha_G$ and declination $\delta_G$ of the shower radiant [deg], J2000, before acceleration by Earth gravity*;
radiant drift in right ascension and declination, in [deg/day], J2000;
-geocentric speed $V_G$ [km$\,$s$^{-1}$] before acceleration by Earth gravity*;
geocentric ecliptic longitude $\lambda_G$ and latitude $\beta_G$ of the shower radiant [deg], J2000;
Sun-centred ecliptic longitude $(\lambda_G-\lambda_S)$ of the geocentric shower radiant [deg];
two \"Opik's angles $\theta, \phi$ describing the direction of the geocentric velocity vector of the meteoroid in the ecliptic Earth-centred reference frame, [deg];
series of flags: a flag denoting the method used to averaging the meteoroid parameters, a flag indicating the known or not known origin of the stream, a flag indicating the credibility assessment (or not) of the stream;
semi-major axis {\it a} [AU];
perihelion distance {\it q} [AU];
eccentricity {\it e};
argument of perihelion ${\omega}$ [deg], J2000;
longitude of ascending node ${\Omega}$ [deg], J2000;
inclination of the orbital plane to the ecliptic {\it i} [deg], J2000;
number of members of the identified stream {\it N* };
name of the parent body, if suggested;
remarks, if any;
observation technique: C-CCD, P-photographic, R-radar, T-TV or video, V-visual*; 
look-up table file name*;
bibliographic reference (if possible, the ADS link or the regular bibliographic code, or the first author and the name of the journal for which the publication is planned)*.\\

The optional parameter $\lambda_G-\lambda_S$  was introduced because of its ability to compensate for the diurnal motion of the meteor's radiant, while the $\theta$ angle was added due to its quasi-invariant nature, (see \citet{1999MNRAS.304..743V} and \citet{1999CeMDA..73...55F}). Both $\lambda_G-\lambda_S$ and $\theta$ make it easier to determine whether there are duplicates of streams in the new data provided to the MDC (see Section \ref{duplicates}).

\subsubsection{Look up tables}
\label{submissionLuT}
From 2019, for each meteor shower, more detailed data are required in the form of the look-up table (LuT) 
when submitting to the MDC. The following information should be provided for each member of the meteoroid stream, (mandatory data are marked with an *):
moment of time corresponding to meteor observation (UTC timescale)*; geocentric right ascension and declination of the meteor radiant [deg], J2000*;
geocentric speed [km$\,$s$^{-1}$] of the meteoroid before acceleration by Earth gravity*;     
ecliptic longitude of the Sun at the meteor instant [deg], J2000*;
Sun-centred ecliptic longitude of the geocentric radiant [deg];
ecliptic latitude of the geocentric radiant [deg], J2000;
code of the source catalogue of the meteor.\\

Details on the required data format of the LuT records are given in the actual template posted on the MDC website.
Experience with the LuTs has shown that it would be reasonable to add other parameters (some obligatory and some optional), such as 
the size of the sample in which the identification was made; the reference to the publication which describes submitting data; the reference to the meteoroid stream identification method; the reference to the name of the parent body. The orbital parameters of submitted shower members would also be highly valuable and appropriate for a direct connection to the orbital part of the MDC. These points, however, require deeper discussion in the IAU F1 WG. 

\section{Established status of MDC showers}
\label{MDCestablished}
\subsection{Difficulties with the 'establishing' process}
As stated in Table \ref{tab:msdlists}, in the MDC SD, meteor showers are grouped into three lists according to their different status. One of these lists is the List of established showers, currently containing $110$ items. 
Giving the status of an established shower means that we are dealing with a real stream and that its parameters (radiant coordinates, velocity, and orbit) have been well determined. Showers with established status receive their names officially during IAU general assemblies.

Preparation of the list of candidate meteor showers for receiving the established status is the main task of the IAU F1 WG \citep[see][]{2007IAUTB..26..140S}.
%
%
However, the WG has never formulated a set of criteria that are strictly necessary for such nominations in order to avoid possible extremely subjective judgement. Hence, the practice of nominating meteor showers for inclusion on the List of established showers has been problematic ever since the beginning of the operation of the MDC database.

The established status of a meteor shower (stream) means that the shower is well-confirmed and there are no doubts about its existence. This can be achieved in several ways, i.e. by repeated observations of the shower in various years, by various techniques, by various teams of authors, by means of an appropriate method of identifying the members of the stream in conjunction with an evaluation of the reliability of the results obtained, and/or by knowing its parent body. 

It is a complicated process and requires deep consideration of each individual shower. However, for the first selection, there was a need for objective rules. We generated a set of such criteria to be met by any stream proposed for the nomination. 
Their final form, provided below, is the result of a discussion that took place among the WG members.

\subsection{Criteria for moving showers to the List of established showers}
\label{C-criteria}

A meteor shower listed in the MDC database can be nominated for established
status by anyone if the shower fulfils criteria C1-C3. Criterion C4 is not obligatory but will strengthen the shower candidates position.

\begin{enumerate}[label=C\arabic*.] 
\item{The shower is represented by at least two sets of parameters (hereafter:
solutions) that have been observed or determined
by at least two independent author teams.
}
\item{Each solution was identified using at least 15 members of the shower (in the case of single-station observations, at least 50 members) within one period of shower activity.
}
\item{The showers existence is supported by an evaluation of its statistical significance in the local sporadic background, or the existence is supported by an estimate of the probability of a random
coincidence of the orbits in the dataset used.
}
\item{The stream parent body is known, or there is a candidate for the parent body, or another
mechanism of the stream formation is suggested. 
}\\
Compliance with the C3 and C4 criteria must be confirmed by research and publication submitted to the MDC.   
\end{enumerate} 

The WG must unanimously approve each nomination. Once the WG has approved a suggestion, the president of the WG, in agreement with the president of the F1 Committee, contacts the IAU.
The nomination proposals are presented at committee meetings during the GA IAU.
Once the IAU has approved a nomination and the shower has officially received its name, it will be moved from the Working list to the List of established showers in the MDC.
The criteria for a stream nomination were approved by a majority of members of the IAU Commission F1 who participated in the electronic vote completed on July 20, 2022.

\section{Streams removed from or moved within the Working list}
\label{MDCremoved}
Another list in the MDC with a special status is the list of data concerning removed showers. This list contains those solutions to a given stream that are subject to some uncertainty.

The transferring of shower data to the List of removed showers requires adhesion to explicit rules to ensure objectivity. 
After publishing its discovery, the meteoroid stream placed on the Working list may remain on this list for many years; in theory, this could be indefinite (see Section \ref{structure}). In practice, a stream can be moved from the Working list to List of removed showers if such a recommendation is published in the literature. Such transfer was approved by the WG in 2019; however, no criteria were given for such a recommendation, and it was agreed that there was no obligation to notify the discoverer of the stream about such action \citep[see][]{2020PSS..18204821J}. In this section, we discuss some of the reasons why such a transfer is possible. We also propose a number of conditions that should be met for this purpose. The proposed criteria have been discussed among the members of the Working Group on Meteor Shower Nomenclature.

\begin{table}[!ht]
\centering
\caption{The MDC data from the List of removed showers consists of 208 records/117 showers.}
\begin{tabular}{lr}
\hline
{ } & {Number}   \\
\hline
{Records without or incorrect references} &  94 \\
{Solutions identified by N < 3 members} &  40 \\
{Duplicates} &  8 \\
{Showers of low reliability} &  55 \\
{Recommended to remove from other reasons} &  11\\
\hline
\end{tabular}

\label{tab:removed}
\end{table}

\subsection{Statistically insignificant showers}
\label{unreliable}
 In the MDC, out of $2000$ data records (solutions), 
 there are $12$ streams identified by $1$ 
 member, $28$ by $2$ members, and $48$ by $3$ members. On the other hand, the most numerous stream consists of $10381$ orbits. The credibility of these streams is certainly not the same. Sixty years ago, streams were searched among photographic samples of several hundred orbits. Currently, the orbital sample sizes are up to several hundred thousand. Therefore, meteoroid streams consisting of a small number of orbits, identified among a large data sample, cannot be considered reliable, especially if their identification was not accompanied by an assessment of their reliability. According to \citet{ 2017PSS..143...43J}, this is particularly the case for streams with fewer than three members. 
 
 In this context, we suggest adding an obligatory parameter that defines the size of the sample in which the shower identification was made to the LuT mentioned in Section \ref{submission}. We also encourage the authors to examine the reliability of 'new' showers. Moreover, the evaluation of a shower's statistical significance in its local sporadic background is one of the new criteria for achieving established status (see Section \ref{C-criteria}). There are various procedures that reflect the strength of a shower compared to its local sporadic background; for example the break-point method developed by \cite{1995EMP...68..427N} (see also the detailed description of the method by \cite{2019msme.book..161V}), the methods introduced by \cite{2016MNRAS.455.4329M} and \cite{Sugar_etal2017}, 
 or the methods suitable in the case of radar data such as the 3D wavelet transform by \citet{Brown_etal2008} or its recent improvement by
 \citet{Kipreos_etal2022}.
Another approach is to estimate the probability of a random coincidence of two orbits, which helps to set a threshold value of the D-discriminant for a specific sample of orbits, and, thus, discriminate between the related orbits and those which are similar by chance (\citet{2017PSS..143...43J} \citet{2020Icar..35113960H}).
 \subsection{Streams with unpublished data sources}
 \label{unpublished}
 Another reason for the lack of reliability in the meteoroid data reported in the MDC is the lack of bibliographical information about the journal in which these data were published. As we mention in Section \ref{procedure}, the publication of the parameters of meteoroid streams is a necessary condition for them to be included in the MDC. In the case of new data sent to the MDC, their authors have six months to submit the relevant publication. In these cases, the lack of bibliographic information is excused. Unfortunately, in the MDC, there are still records of data for which we do not have original bibliographic information. We have discovered $94$ such cases. All of them relate to meteor data provided to the MDC prior to the introduction of the requirement for an original publication describing the meteor shower. Other showers of the first $230$ streams of the MDC SD (see Section \ref{MDCbegi}) have been verified using the original source publication or the bibliographic data provided by \citet{2006mspc.book.....J}. 
 %
\subsection{Duplicates}
\label{duplicates}
A new stream sent to the MDC may turn out to be a duplicate of a stream already existing in the SD. This is judged on the basis of the clear similarity between the parameters of the new stream and one of those already existing in the database. 
In the case of duplicates found (published and sent to the MDC), the 'new' stream is treated as another solution of the 
previously known stream. The SD team will inform the author(s) that their shower data has been moved to the List of removed showers (under their 'new' shower name) and to a previously discovered shower as its additional solution.

However, discovering duplicates is no simple task. A comparison of the averaged values of the orbital parameters of two showers, by calculating the value of a D-function (\citet{1963SCoA....7..261S}, \citet{1981Icar...45..545D}, and \citet{1993Icar..106..603J}), would only indicate a possible similarity and would lack consistency. The situation is more promising with showers for which the parameters of individual shower members are known. In the MDC SD, there are parameters of individual members for only about 10\% of meteor showers. The obligation to submit them in the form of the LuT was established at the June 20, 2019 meeting of the WG at the 'Meteoroids 2019' conference (see Section \ref{submissionLuT}). LuTs 
contain parameters that describe only the dispersion of a showers solar longitude and the radiant and speed for each member of a shower \citep{2020PSS..18204821J}. Regarding older streams that were submitted to the MDC before 2019, these parameters are missing. 

We would like to emphasise here that it is not the task of database operators to decide whether a new submitted stream is a duplicate of a stream already present in the MDC. It is the responsibility of the reviewer of the publication describing the new shower discovery, or anyone who undertook such evaluation and published it in a peer-reviewed scientific journal or one of the amateur journals: the WGN (Journal of the IMO) or Meteor News. However, for authors submitting new showers to the MDC, we are preparing a simple tool for an approximate check of a new shower in relation to the others in the database (see Section \ref{MDCnew}). Yet, to accomplish the process of a consistent search for duplicates, deeper analyses will be needed. 

\subsection{False-positive duplicates}
\label{identity}
The opposite problem to duplicates is the problem of the identity of a shower. Showers with problematic identities are those whose various solutions are not compatible and which should have been submitted to the MDC SD as different showers. Such showers are also present among the established ones. Solving this problem will be the task for a near future work (see also Section \ref{MDCverification}). Moving such showers back to the Working list seems to be necessary; however, it requires a discussion within the WG. 
%
\subsection{Criteria for moving showers to the List of removed showers}
\label{R-criteria}
In conclusion, given the above, we present the proposed criteria for a removal.
A meteor shower (meteoroid stream) listed in the MDC SD shall be moved to the List of removed showers if one of the following criteria applies:

\begin{enumerate}[label=R\arabic*.]
\item {The correct bibliographic information for the stream identification is absent.}
\item{The shower is found to be a duplicate (see Section \ref{duplicates}) of an earlier discovered shower.}
\item {The shower has been found to be unreliable (see Section \ref{unreliable})}.
\item {The stream was identified using fewer than three meteoroids.} 
\end{enumerate}

Compliance with the R2 and R3 criteria must be confirmed by research and publications submitted to the MDC.
Moving a shower to the List of removed showers does not mean its elimination from the MDC SD. Each shower on that list has the possibility of returning to the Working list when the criterion according to which it was removed (confirmed by research, an article published and sent to the MDC) no longer applies.
The suggested criteria for removing showers were approved by a majority of members of the IAU Commission F1 who participated in the electronic voting completed on July 20, 2022.

In some cases of showers with multiple solutions, not all solutions need be removed from the Working list, and the above described criteria are applied only to one or more individual solutions. The current status (October 2022) of the List of removed showers is shown in Table \ref{tab:removed}.
%
\section{Verification of the MDC SD}
\label{MDCverification}
In this chapter, we describe the implementation of the first step of the MDC data verification process. We are working on the next steps (see Section \ref{furthersteps}), which will be presented in future papers. 

In $2007$ the data of $\sim$$230$ meteoroid streams, mainly from Table $7$ in \citet{2006mspc.book.....J}, were selected for the MDC SD. Subsequent portions of meteor data came via e-mail submissions by their authors.

The impetus to modify and verify the MDC database was the critical remarks on the content of the database and on its functioning expressed within the meteor community, by professionals as well as amateurs (see \citet{2014pim4.conf..126A}, \citet{2016JIMO...44..151K}, \citet{2020eMetN...5...93K}, and \citet{2018eMetN...3....1K}). Moreover, the content of the SD has not yet been checked for the correctness of the meteor data contained therein.

\subsection{Implemented modifications }

An obvious way to verify the correctness of the data contained in the MDC is to compare them with the corresponding data provided in the source publications. Although in the era of electronic data transfer, it might seem unnecessary, in fact it turned out to be the most justified. We quite often found that the data provided electronically to the MDC differed slightly from those contained in the publication.
Except when we could not obtain access to the publication, we made such a comparison. 
We have adopted a rule to restore the values as they are given in the publications to the database, without any processing. Hence, in the course of this work, if necessary, we restored the published values of all parameters of the streams and, at the same time, supplemented the database with some missing parameters and information. All the changes explained in the following paragraphs are summarised in Table \ref{tab:verification}.
%
%

When it was possible, instead of a general annual shower activity, we provided the years in which the stream members were observed. We also added the ecliptic longitude of the Sun at the beginning and end of the shower observation ($\lambda_{Sb}, \lambda_{Se}$).
We restored all angular parameters to values corresponding to the epochs given in the source publications. And in each of these cases, an appropriate flag was introduced. 
Similarly, we restored the meteoroid velocity values given in the publications, and when they were out-atmospheric velocities ($V_{\infty}$), the appropriate flags were set up. We introduced the missing values of the eccentricities of the stream orbits.
A few more flags were introduced, telling of: when the median values of the radiant and the orbital parameters are given; when the method used for assessing the reliability of the stream is known; and when the parent body of the stream is given in the publication. 
Finally, we introduced the date of the shower submission into the MDC SD, which is vital information for determining the discoverer of a particular stream.

 Wherever we considered it advisable, we added page numbers and/or tables indicating where the shower can be found in the original publications. Also, we add a comment if the name of the shower given in the MDC is not the same as in the publication (labelled as 'Records with changed names' in Table \ref{tab:verification}). 

Currently (October 2022), the MDC SD contains $1378$ data records (solutions) for $923$ showers and $208$  meteor shower data records with the 'removed' status (Table \ref{tab:verification}). For $194$ streams, two or more data records are available. We made $1732$ modifications, some of which consisted of introducing new or correcting existing data. 

Over the last few months, we have obtained new submissions in the MDC SD, the papers of which have not yet been published. These records will be continuously checked after their publication.

When verifying the MDC data, $94$ records with problematic or incorrect references were found. These include cases for which relevant shower data cannot be found in the given reference or no radiant data are specified in it, or where the reference is unknown or not accessible.
Furthermore, the information on the $N$ -- number of identified
stream members was neither found in the submissions to the MDC nor given
in the original publication $141$ times. 

Within the verification process, we found $74$ showers that met the criteria for removal. The number of removed showers increased from the original $43$ to $117$ (see Table \ref{tab:removed}). Two showers among them are from the List of established showers: 102/ACE and 252/ALY. It is necessary to discuss moving their solutions back to the Working list with the WG members and to notify the IAU Commission F1.

\begin{table}[!ht]
\centering
\caption{Summary of the verification process of the MDC data from the List of all showers.}
\begin{tabular}{lr}
\hline
{ } & {Number}   \\
\hline
{Records possible to verify} &  1378 \\
{Verified records} &  1364 \\
{Still to be checked} &  14 \\
{Corrected records} &  1212 \\
{Corrections made} &  1732 \\
{$\lambda_{Sb}, \lambda_{Se}$ added } &  659 \\
{Records with changed names} &  101 \\
{Records with Bessel epoch $B1950$} &  215 \\
{Records with $V_{inf}$} &  140\\
{Records with median} &  161\\
\hline
\end{tabular}
\label{tab:verification}
\end{table}

\subsection{Discussion of additional issues}

Here we discuss some of the issues which were addressed to the MDC by database users but could not be fully implemented or which need an explanation. They involve incompleteness of shower data, the inconsistency of shower parameters, and the different types of shower parameters used in the MDC SD.   

\subsubsection{Incomplete shower data}
With regard to 
Incomplete shower data, for example 
missing orbital elements, we note 
that the MDC SD is not intended to include all information on meteor showers; the only mandatory parameters required are the years of the shower observations, the averaged value of the ecliptic longitudes of the Sun corresponding to the moments of observations of meteors, the averaged geocentric radiant coordinates and velocities, and the number of members of the identified shower. By 'averaged' values we mean values calculated as an ordinary arithmetic means of individual parameters or averaged using the method given by \citet{2006MNRAS.371.1367J, 2008MNRAS.384.1741J, 2010pim8.conf...91J} or as median values of individual parameters.

We strongly encourage those who submit new data to the MDC to provide, in addition to the mandatory data, a set of averaged orbital elements, an assessment of the reliability of the stream identification as well as potential parent body of the meteoroid stream.

\subsubsection{Inconsistencies among the shower parameters}

Several users of the MDC SD noted a certain inconsistency between the given mean values of the Sun's ecliptic longitude and the longitude of the ascending node of the averaged orbit. Usually, the differences in these values are close to $0$ or $180$ degrees, depending on what node in the meteoroid's orbit the Earth was at the time of meteor observation. In fact, some level of inconsistency between the averaged values of the radiant and the averaged orbit elements is understandable, for example among the individually averaged $a, q,$ and $e$ values. However, this is only up to a certain level.  We analysed each suspicious case separately during the MDC meteoroid data verification process.

It is worth recalling here that, even in the case of an orbit of a single meteoroid, the difference between the ecliptic length of the Sun at the time of meteor observation and the length of the ascending node of the orbit, need not necessarily be $0$ or $180$ degrees. In the catalogues of meteoroid orbits, one can find such orbits for which these differences are sometimes significantly different. A meteoroid moving in an orbit with a slight inclination to the ecliptic plane need not be observed in one of its orbit nodes (see \citet{2010pim8.conf...91J} and \citet{2010epsc.conf..888J}).

\subsubsection{Inconsistent type of shower data within the database}
For some MDC users, it may be unclear why the values of the angular parameters of the radiants and the orbital elements are given in different epochs, sometimes B1950, sometimes J2000; why different type of velocities of meteoroids ($ V_G$ and $V_{\infty}$) are given; or why different methods of averaging the values of these quantities (arithmetic means, medians, etc.) have been used. This results from the principle which we have adopted, namely that we enter data into the MDC as they are given in the original publication
However, the user of the MDC database now has the possibility to select between the original data and those converted into a common epoch and type of geocentric velocity. The method of averaging the values of the parameters remains unchanged.

\section{New version of the MDC SD: Contents, website interface, and tools}
\label{MDCnew}
\subsection{Standardisation of the database content: Additional parameters}
Previously, the meteoroid data collected and presented on the MDC website were not uniform. The angular values of the parameters were expressed in different reference frames. Earlier data were given in the B1950 epoch and later ones in the J2000 epoch. Also, the geocentric velocities of the meteoroids could correspond to values that were corrected (or not) for the gravitational pull of the Earth. The user of the MDC database had no possibility (apart from checking in original publications) of determining what epochs and speeds they were dealing with. 

In the current version, these issues have been unified. All angular data available in the MDC database are expressed in the J2000 reference frame. Data from the B1950 epoch were transformed using the expressions given in \citet{1992esta.book.....S}. All extra-atmospheric velocities $V_{\infty}$ were corrected for the gravitational pull of the Earth by the formula taken from \citet{1987BAICz..38..222C}. 

As mentioned in Section \ref{submission}, five additional (optional) parameters were entered in the meteoroid data record: the ecliptic longitude $\lambda_G$ and latitude $\beta_G$ of the shower radiant as well as the sun-centred longitude $\lambda_G-\lambda_S$ of the radiant. We also added two angles, $\theta$ and $\phi$, which together with the geocentric velocity $V_G$ constitute a triplet of {\"O}pik variables, which we have borrowed from his theory of close encounters described in \citet{1976iecr.book.....O} and \citet{1990CeMDA..49..111C}. 
Due to the quasi-invariant nature of $V_G$ and $\theta$ (see \citet{1999MNRAS.304..743V} and \citet{1999CeMDA..73...55F}), together with the sun-centred longitude, these quantities make it easier to determine whether there are duplicates of streams in the new data provided to the MDC. 

\begin{figure*}[h]
\centerline{
            \includegraphics[width=0.20\textwidth, trim=2mm 2mm 2mm 2mm, clip]{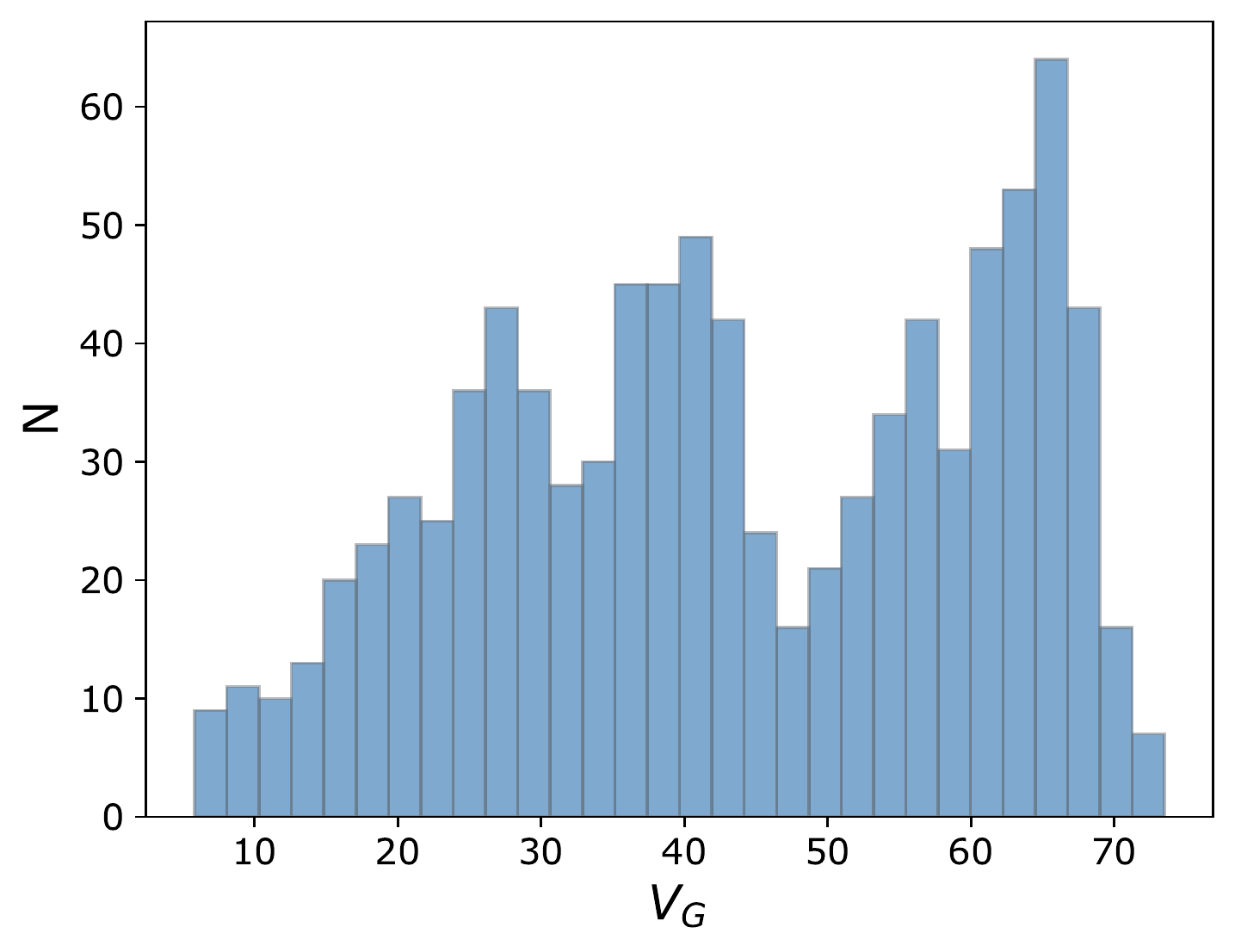}
            \includegraphics[width=0.20\textwidth, trim=2mm 2mm 2mm 2mm, clip]{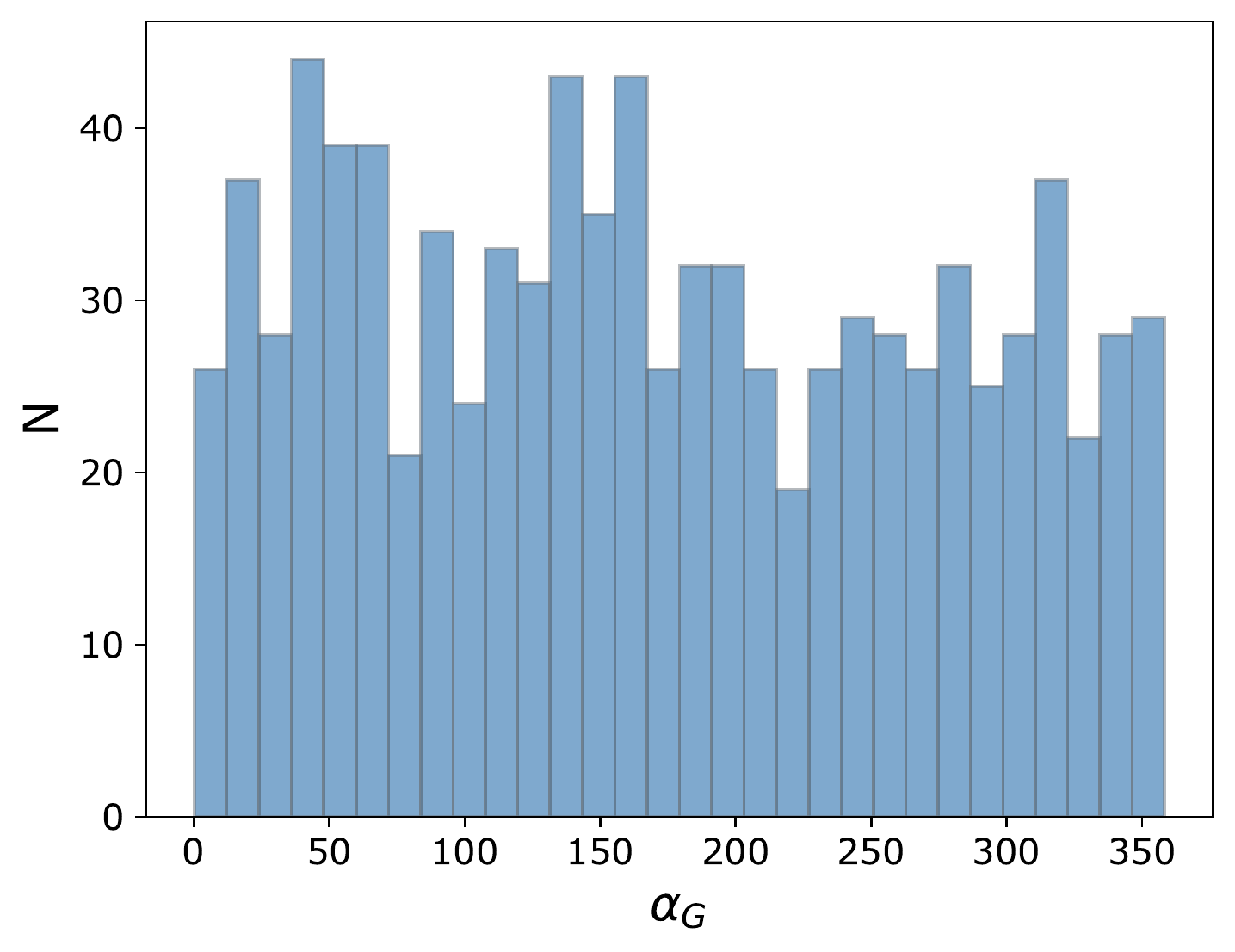}
            \includegraphics[width=0.20\textwidth, trim=2mm 2mm 2mm 2mm, clip]{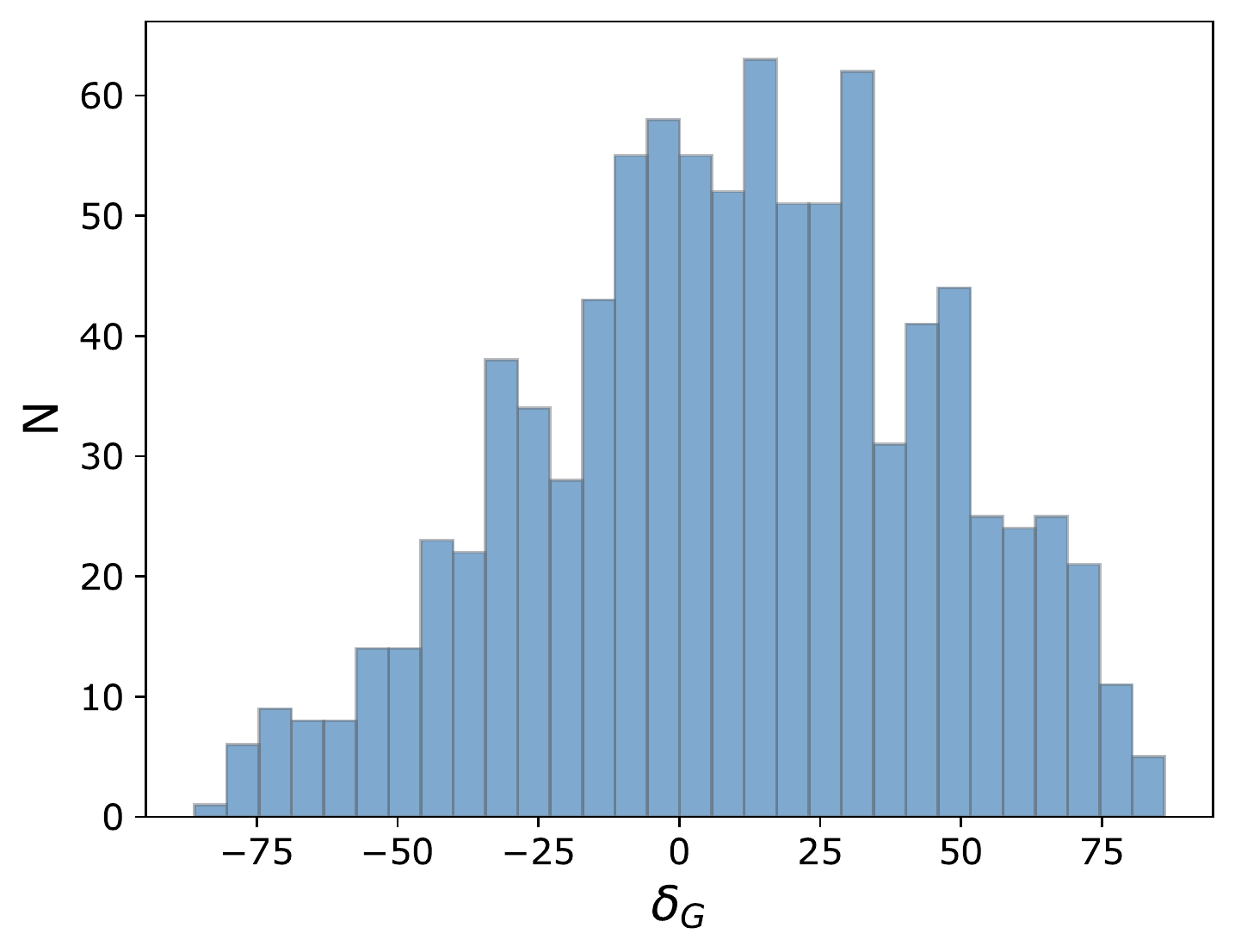}
            \includegraphics[width=0.20\textwidth, trim=2mm 2mm 2mm 2mm, clip]{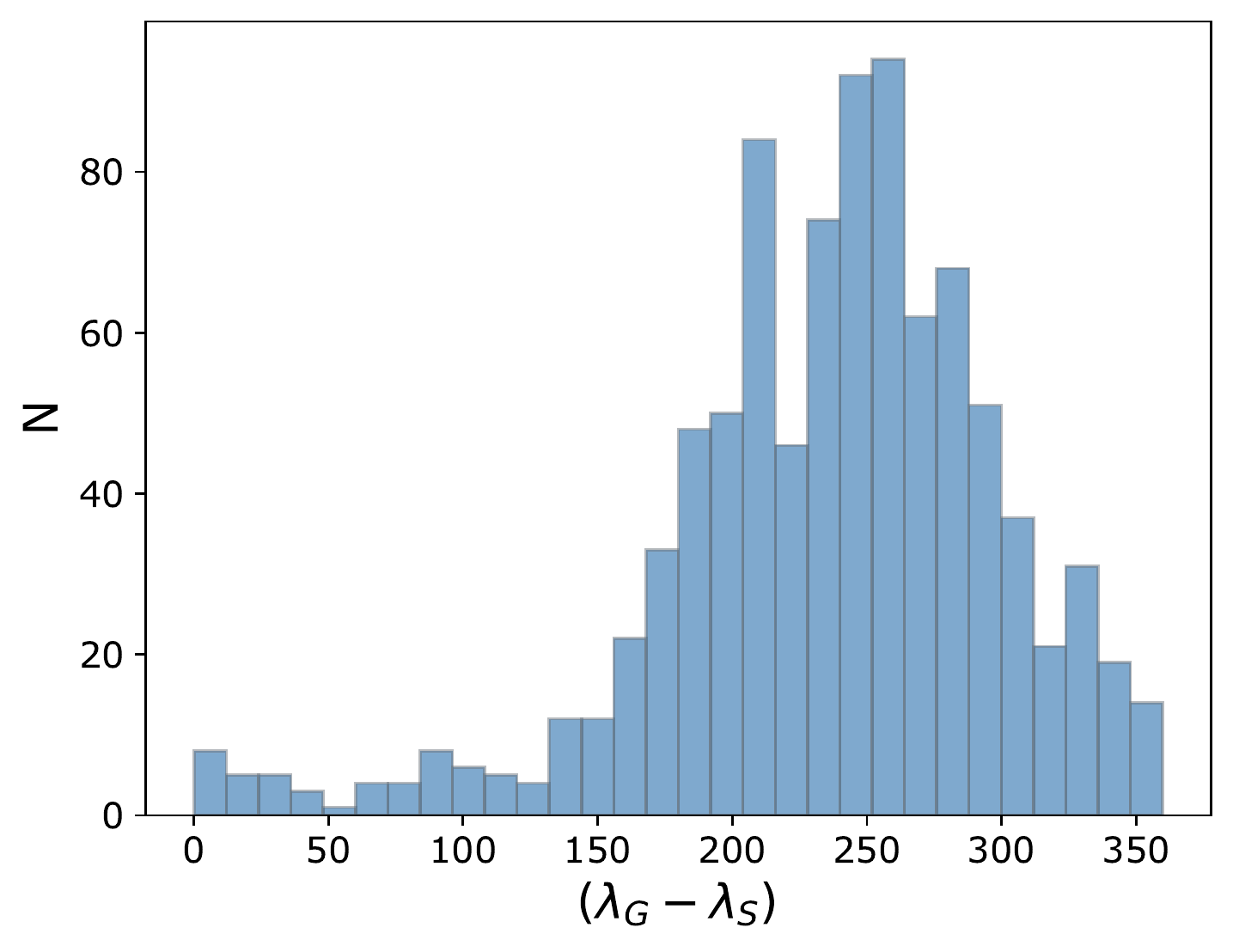}
            \includegraphics[width=0.20\textwidth, trim=2mm 2mm 2mm 2mm, clip]{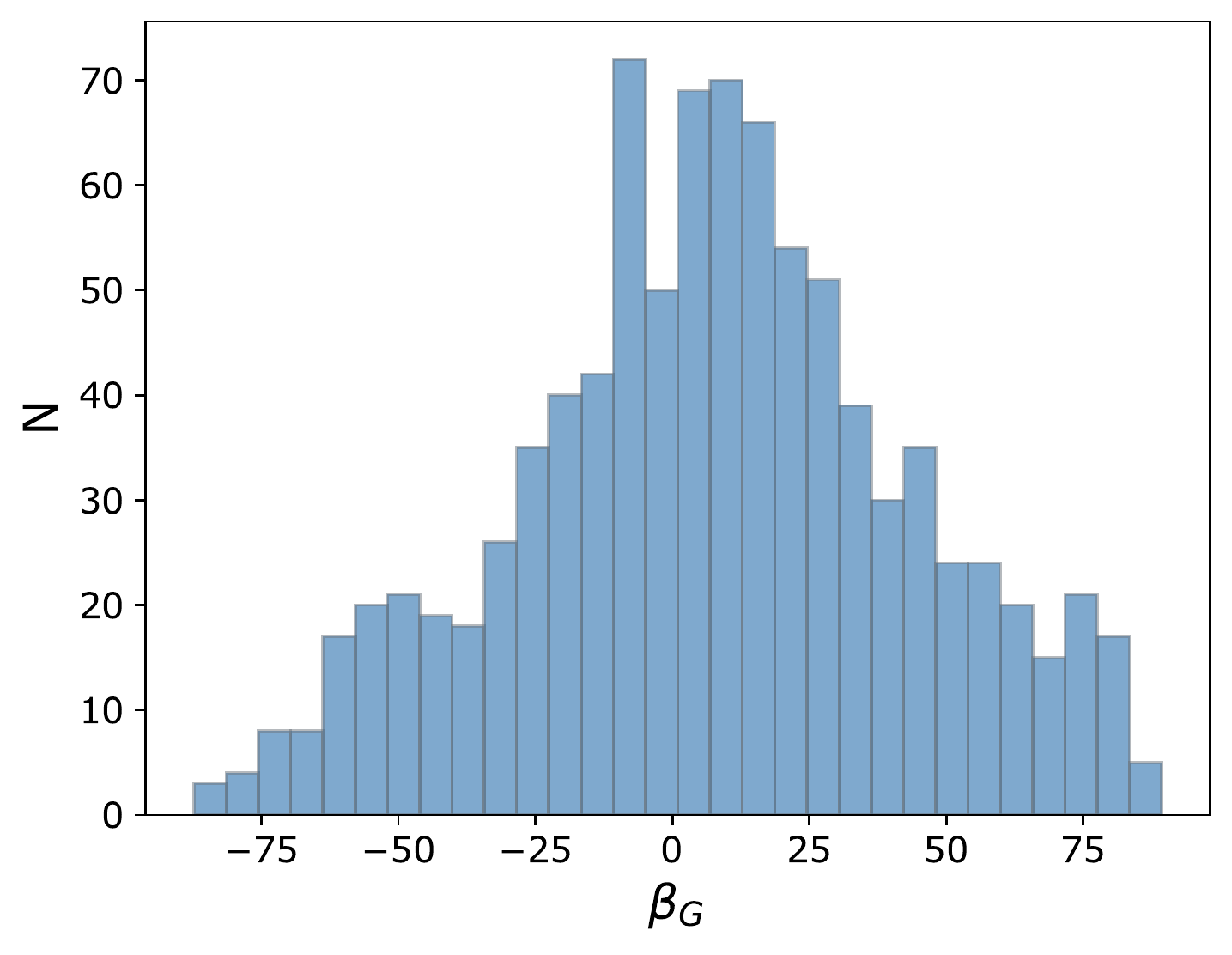}            
            }
\centerline{
            \includegraphics[width=0.20\textwidth, trim=2mm 2mm 2mm 2mm, clip]{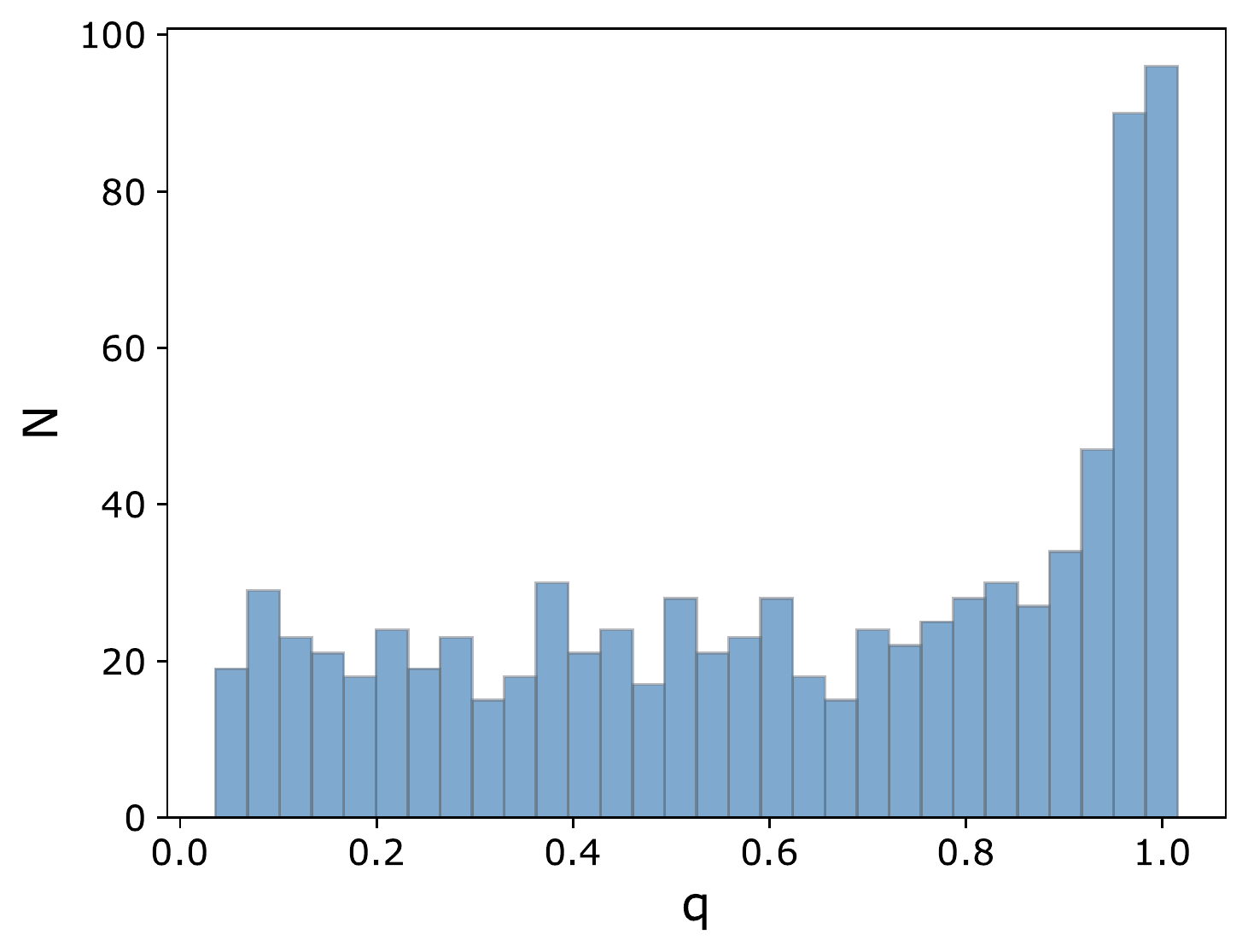}
            \includegraphics[width=0.20\textwidth, trim=2mm 2mm 2mm 2mm, clip]{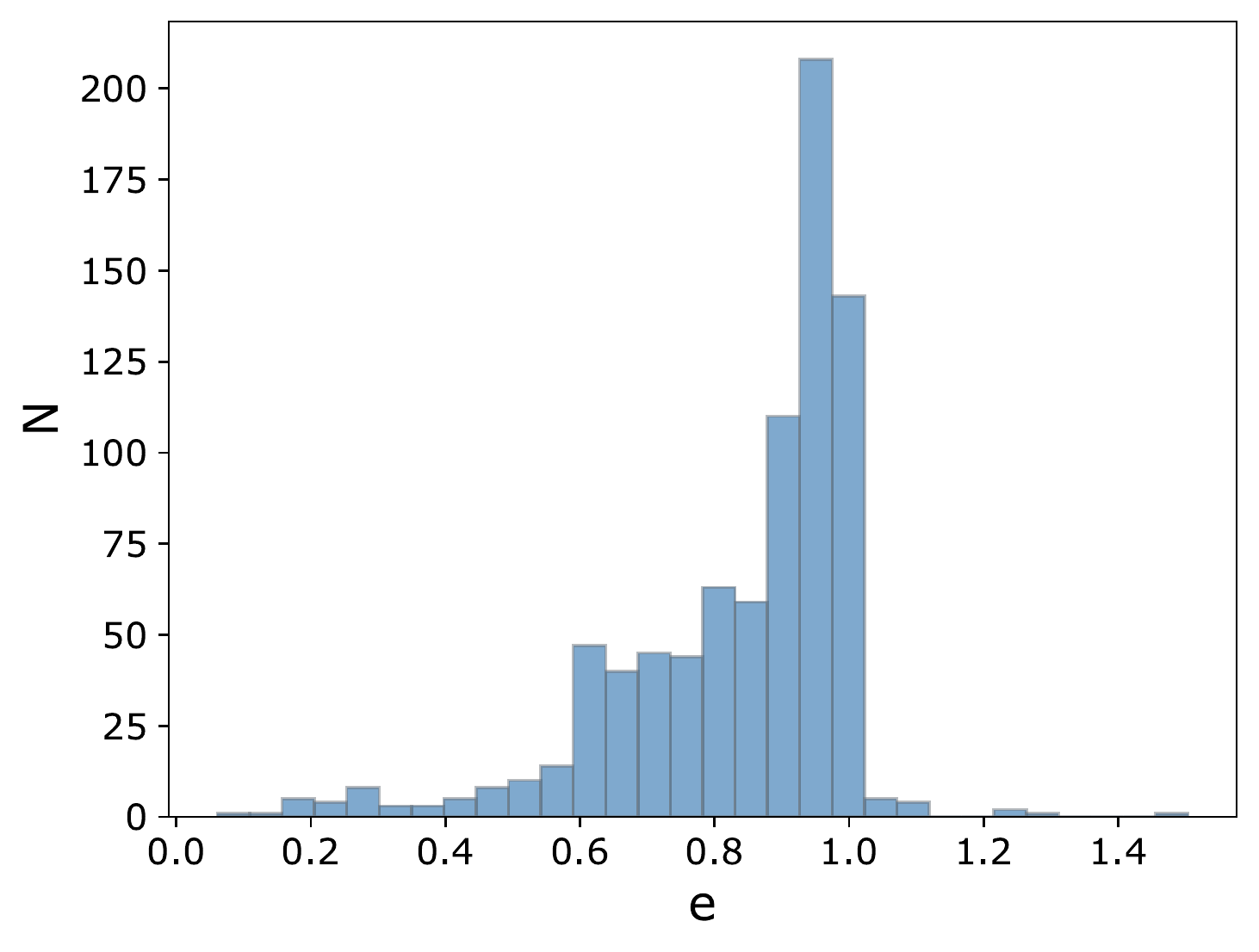}
            \includegraphics[width=0.20\textwidth, trim=2mm 2mm 2mm 2mm, clip]{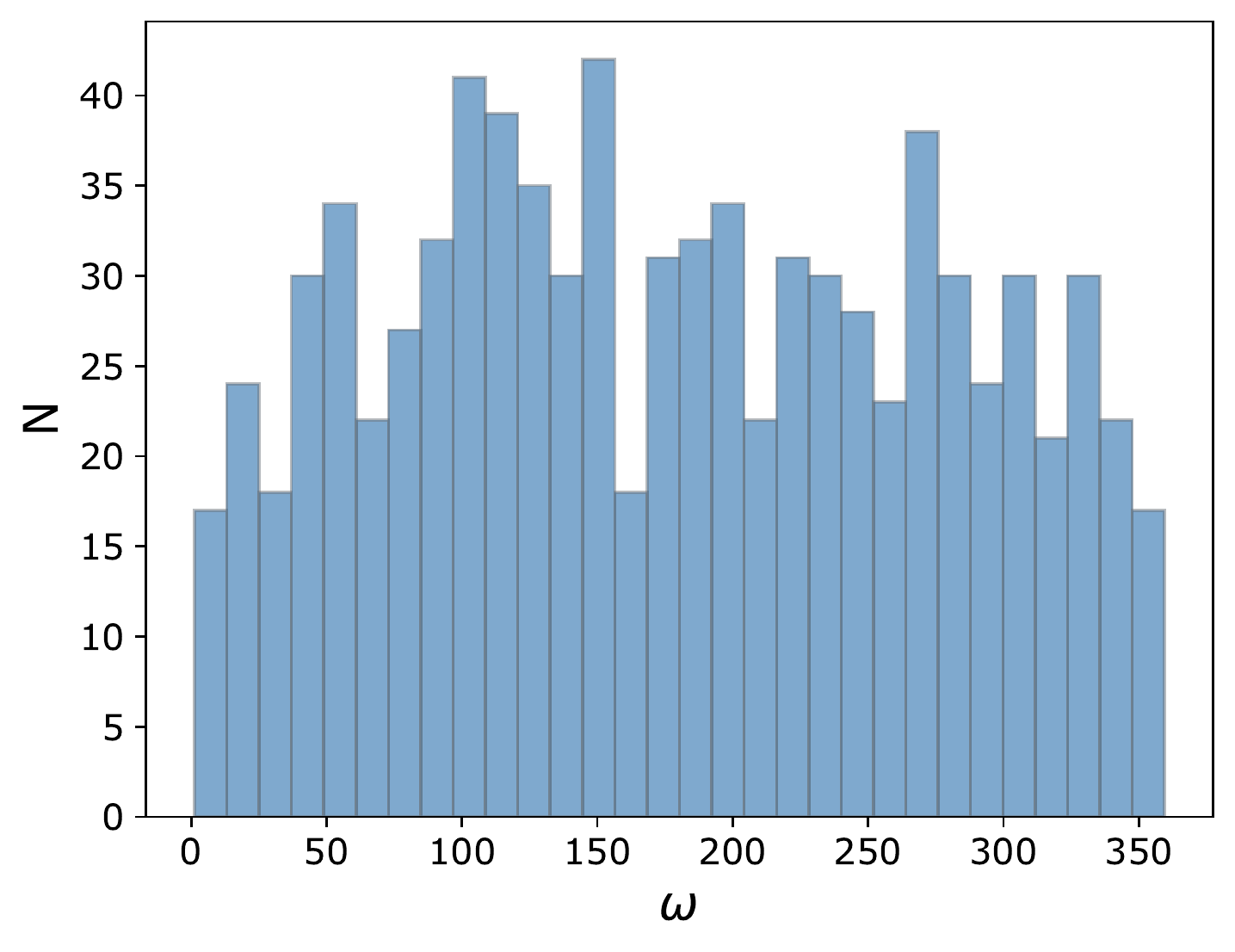}
            \includegraphics[width=0.20\textwidth, trim=2mm 2mm 2mm 2mm, clip]{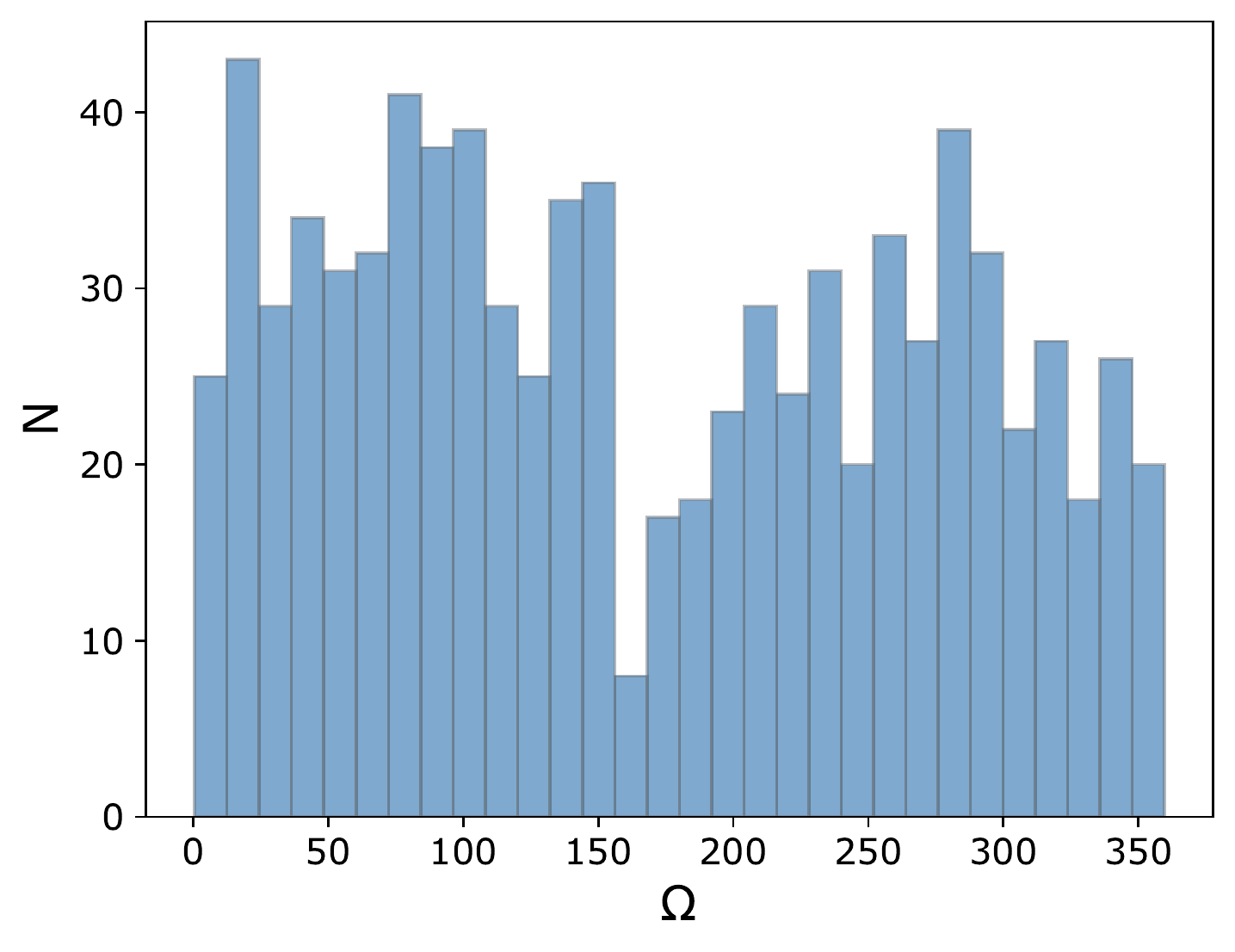}
            \includegraphics[width=0.20\textwidth, trim=2mm 2mm 2mm 2mm, clip]{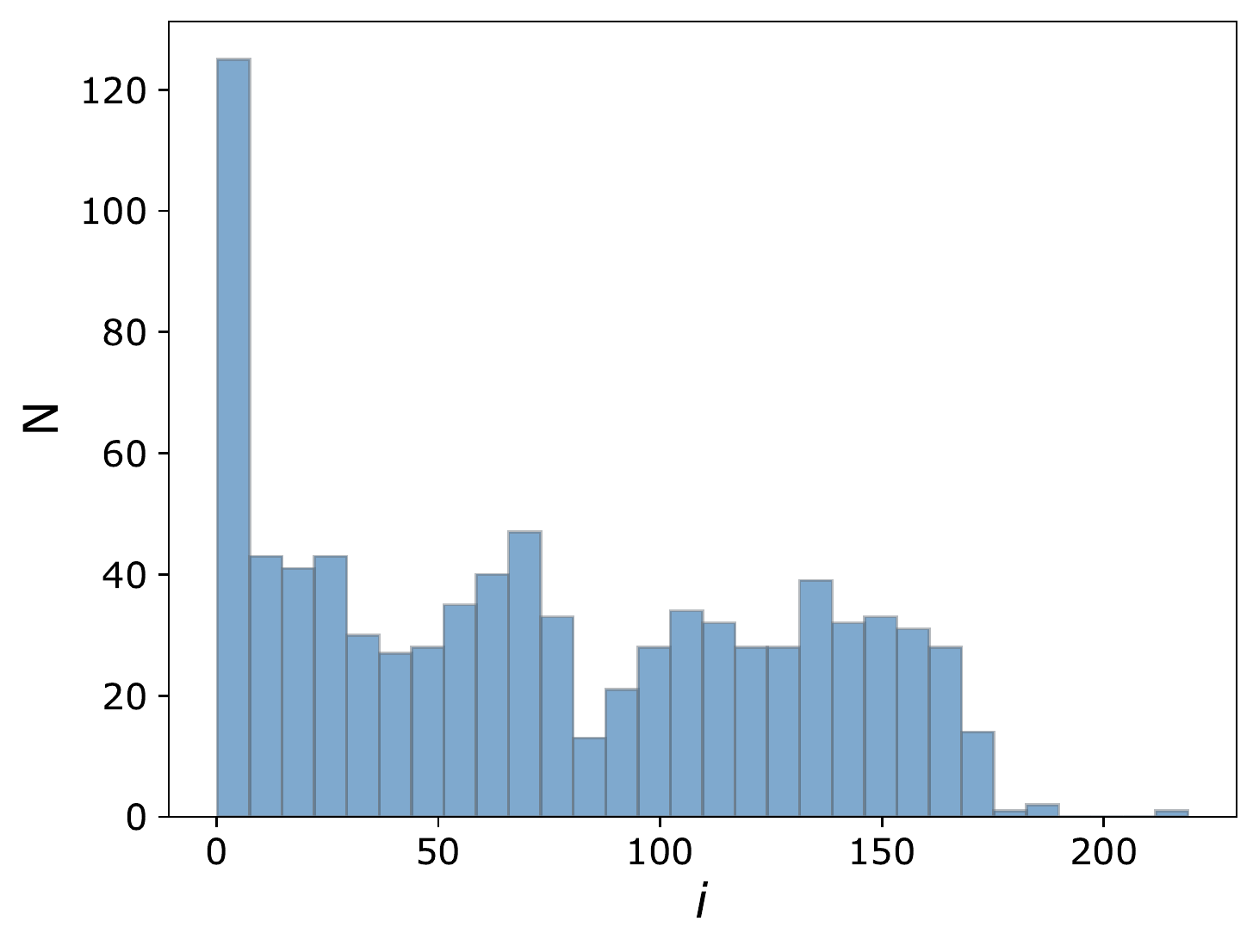}
            }
\caption[f2]{Distributions of geocentric and orbital parameters of all $923$ meteor showers from the MDC: the averaged values of the geocentric velocities ($V_G$), the geocentric right ascension ($\alpha$) and declination ($\delta$) of the shower radiants, the sun-centred ecliptic longitudes ($\lambda_G-\lambda_S$) and ecliptic latitudes ($\beta_G$) of the shower radiants, 
 the perihelion distance ($q$), eccentricity ($e$), argument of perihelion ($\omega$) and the longitude of the ascending node ($\Omega$), and inclination of the orbital plane ($i$).
}
\label{geoparameters}
\end{figure*}

\begin{figure*}[h]
\centerline{
           \includegraphics[width=0.35\textwidth, angle=-90, trim=2mm 2mm 2mm 2mm, clip]{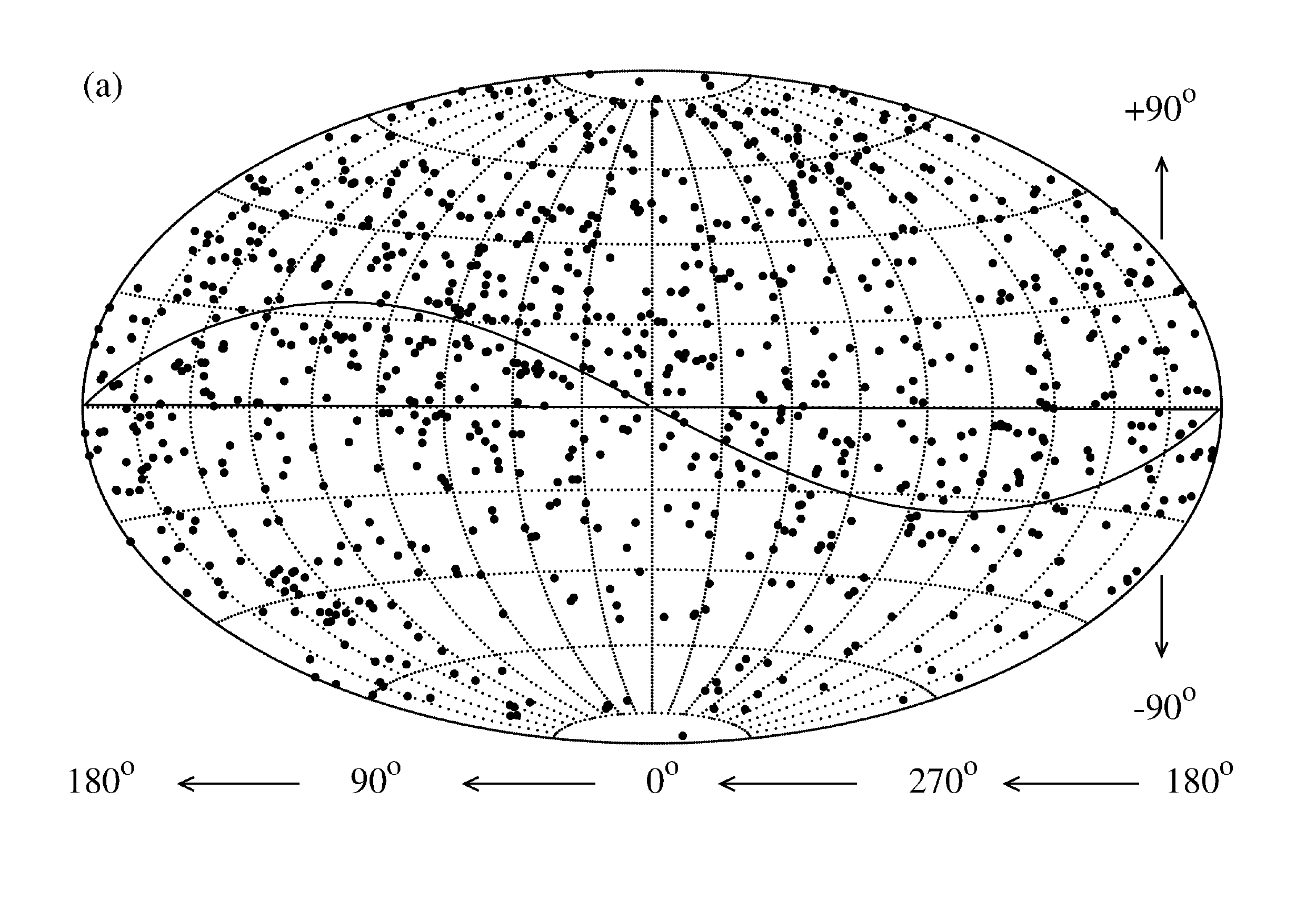}
            \includegraphics[width=0.35\textwidth,  angle=-90,trim=2mm 2mm 2mm 2mm, clip]{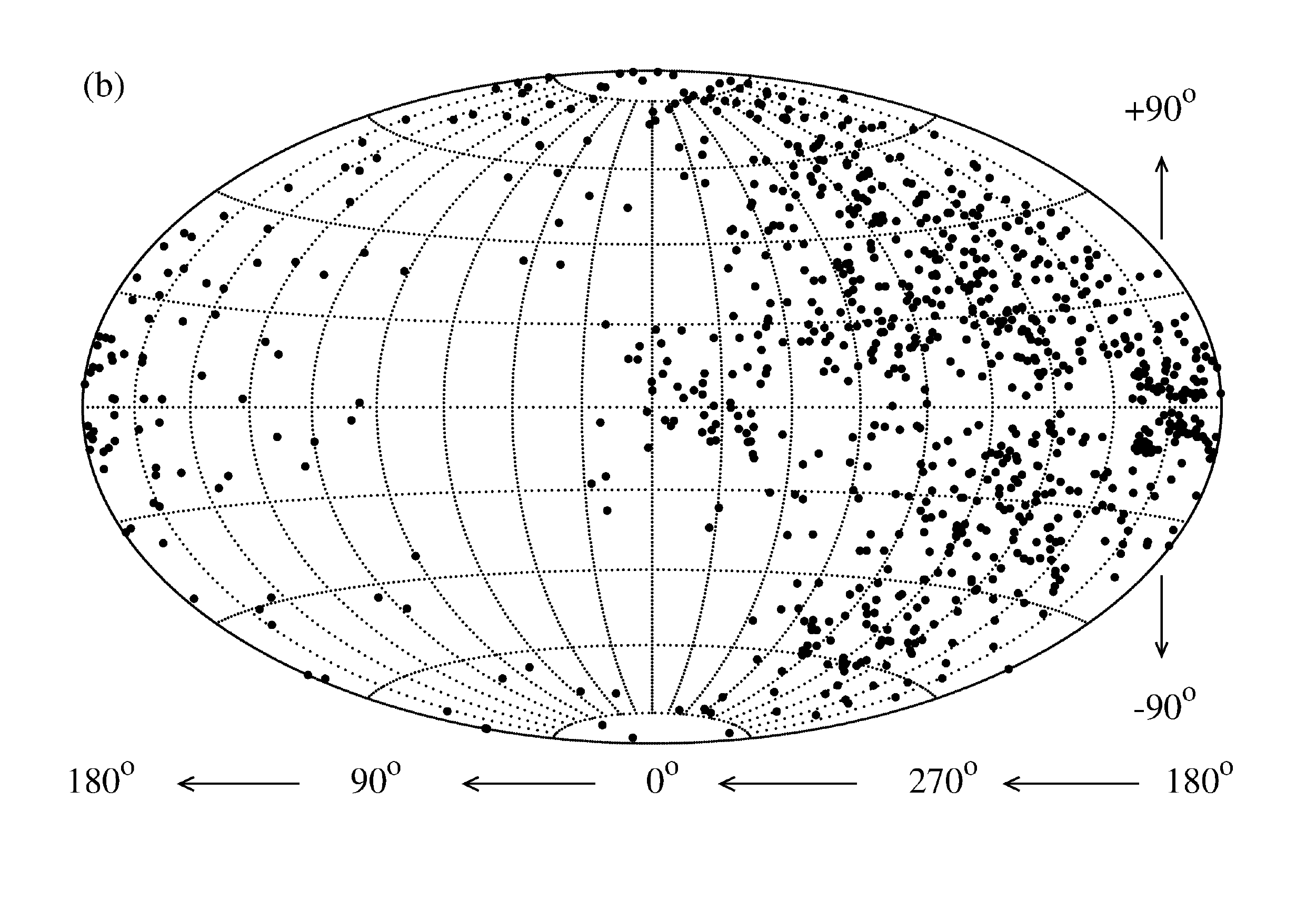}}
\caption[f3]{Hammer-Aitoff diagrams of geocentric parameters of $923$ meteor showers from MDC: (a) distribution of right ascension and declination  of the shower radiants; (b) in the sun-centred ecliptic system, the figure shows the distribution of the ecliptic longitude and latitude of the shower radiants. 
}
\label{hammerdiagram}
\end{figure*}

In the current version of the MDC SD, we have calculated the values of these five additional parameters based on the information available in the shower data provided so far.
Using the equatorial coordinates of the radian $\alpha_G$, $\delta_G$ and the solar longitude $\lambda_S$ at the moment of the meteor observation, the corresponding ecliptic coordinates, $\lambda_G$, $\beta_G,$ and $\lambda_G-\lambda_S,$ were calculated. While using the formulas given in  \citet[][]{1999MNRAS.304..743V} and \citet[][]{1999CeMDA..73...55F}, the angles $\theta$ and $\phi$ were calculated. 
In Figure \ref{geoparameters}, we illustrate the distributions of selected geocentric and orbital parameters  of $923$ meteoroid streams included in the current version of the MDC database. Figure \ref{hammerdiagram} illustrates the $923$ showers radiants in the equatorial and Sun-centred ecliptic systems. 
The distributions of the average values of the meteoroid stream parameters are similar to the distributions of the parameters of the sporadic component. 
 
It may be surprising that, of the $923$ streams listed in the MDC, $46$ have open orbits (Figure \ref{geoparameters}). For $15$ cases, the values of the eccentricities of their parabolic orbits result from the rounding of the calculated more precise values. Thus, there are $31$ streams in the MDC, the mean orbits of which have been determined as hyperbolic. Six of them were determined by radar techniques, the remaining $25$ by video or visual methods.

There are several issues contributing to this problem. Since the heliocentric velocities of the meteoroids of these streams are too close to the parabolic limit, even a small measurement or determination error can create an artificial hyperbolic orbit (\citet{2020PSS..19205060H} and \citet{2020PSS..19004965H}). Another problem is the determination of the mean values.
The orbit with the maximum mean value of the eccentricity $e$$=$$1.504$ was determined by radar technique (shower \#$396$, December theta-Aurigids, \citet{2010Icar..207...66B}). In the case of video techniques, the maximum average eccentricity was $e$$=$$1.265$ (shower \#$691$ zeta-Cetids, \citet{2016Icar..266..384J}). If we test the compatibility of the values of the orbital elements $e, a,$ and $q$, for these two orbits, it appears that only in the case of the December theta-Aurigids is the consistency satisfactory. For the zeta-Cetids, the values of $e$$=$$1.265$ cannot be obtained from the values of $a$$=$$-0.44$ and $q$$=$$0.989$ given in \citet{2016Icar..266..384J}. It turns out that the reason for this situation may be the different methods of averaging the parameters used in both works. In the work by \citet{2016Icar..266..384J}, the median values of individual elements of the orbits were given as average values. In the work by \citet{2010Icar..207...66B}, the mean orbits were computed using the radiant and velocity observed at the time of the maximum of the shower activity. In our opinion, representing the stream with median values taken from individual parameters can be risky. However, we cannot be sure whether the approach used by \citet{2010Icar..207...66B} leads to the correct conclusion in each case.

The issue discussed above shows that, based on the average parameters of the streams given in the MDC, it is not always possible to come to reliable conclusions. We would be in a much better position if the streams in the MDC were represented not by an average orbit but by the orbits of all identified stream members.
\subsection{Modification of the MDC website interface}
In addition to working on the content of the SD, we have been improving the user interface based on present web technologies that enhance the user experience, all while continuing to provide easy access to the contents of the database in its first release. Gradually, we plan to build on the set of tools. For example, authors submitting a shower will have the option of using a simple tool for a quick approximate check of the similarity of their shower to those in the database. Such a tool -- albeit only roughly -- will help authors decide whether the stream they sent does not exist in the MDC or is another solution of an already known stream and should therefore be submitted as such.

We are also preparing to expand the information on the parent bodies of streams listed in the MDC SD by adding a separate reference to the paper in which the research was presented. Currently, in the case of a shower that has several solutions, the note about the parent body does not state in which paper the parent body was proposed. Moreover, there will be the possibility of submitting a suggestion of a parent body (of a stream listed in the SD) that is confirmed by research without submitting the corresponding shower parameters. This option is needed since a known parent body is within the new criteria for achieving established status (see Section \ref{C-criteria}).

\section{Summary and conclusions}
The present paper describes all the modifications (including new procedures) that the MDC management team suggested to, presented to, and discussed with the meteor science community and implemented. The modifications include new information added to shower parameters, which was necessary to make them unambiguous. New procedures for changing the status of meteor showers were necessary to achieve objectivity about such decisions. \\

The summary of the changes in both the maintenance procedures
and the content of the SD are as follows:

(1) We proposed a procedure for the nomination of showers for established status consisting of a set of criteria. The obligatory criterion is based on setting lower limits for the number of several parameters. A candidate stream has to have at least 15 stream members in each of at least 2 solutions which have to be observed by different author teams. The statistical significance of the showers existence should be evaluated. The optional criterion concerns a known parent body.
The application of the stream nomination criteria to the MDC has shown that they are more stringent than the approach used in the past. We believe that the new approach is more transparent.

(2) We have formalised the rules for shifting the shower data from the Working list to the List of removed showers. This transfer will apply to showers: the bibliographic information of which are absent; which have been found to be a duplicate; which were identified using less than 3 meteors; which have been found to be unreliable; or if the author asks to remove them.

(3) We have verified $1364$ meteor data records stored in the SD; $1732$ modifications have been made. Presently (October 2022), the MDC SD data consists of $923$ meteor showers, $110$ are officially named by the IAU; $813$ streams are on the Working List, while $117$ showers meeting the relevant criteria have been moved to the List of removed showers. The list of shower groups has been excluded from the database.

(4) The MDC website has also been modified. The shower part and the section related to individual meteoroid orbits have been combined into a combined website. The shower part is going to be equipped with various tools to facilitate its use in the near future.\\

We cannot claim that the verification and modification of the MDC SD described here has freed it of all defects and that the proposed new rules for determining the status of meteor showers and the rules for providing new data exhaust all the required changes that should be made to the MDC SD. Though the first necessary step has been made and a large number of errors have been eliminated. The procedures used have been freed from their previous subjectivity, and their use may suggest further improvements.

\section{Further steps}
\label{furthersteps}
There are still some open problems (already addressed by us), which, however, need an individual and deep analysis, such as the problem of duplicates and/or false-positive duplicate shower solutions (see Sections \ref{duplicates} and \ref{identity}), difficulties with northern and southern branches of the same meteoroid stream, and questionable information on possible parent bodies listed in the database. Thus, the changes and corrected errors in the MDC database presented in this study do not exhaust the need for further work to improve the database. Therefore, the authors of this paper would appreciate all feedback on the modified version of the MDC SD, as well as on any other issues related to it not raised by us.
\section{Acknowledgements}
The authors like to acknowledge: \v{Z}eljko Andrei\'{c}, Reiner Arlt, David Asher, Rhiannon Blaauw, Ji\v{r}\'{i} Borovi\v{c}ka, Alfredo Dal'Ava J\'{u}nior, Steve Hutcheon, Diego Janches, Peter Jenniskens, Zuzana Ka\v{n}uchov\'{a}, Sirko Molau, Allan Mulof, Francisco Oca\~{n}a, Paul Roggemans, Galina Ryabova, Mikiya Sato, Damir \v{S}egon, Ivan Sergey, Ned Smith and Denis Vida, who in recent years helped improve the IAU MDC contents. 
The work was, in part, supported by the VEGA, the Slovak Grant Agency for Science, grant No. 2/0009/22.

This research has made use of NASA's Astrophysics Data System Bibliographic Services.
\bibliographystyle{unsrt}
\bibliography{IAUMDC_paper_I_language_changes_Arxiv}  
\end{document}